\documentclass[11pt]{article}
\usepackage{graphicx}
\usepackage[margin=1.25in]{geometry}
\usepackage[usenames,dvipsnames]{color}
\usepackage{url}
\usepackage[colorlinks = true,
            linkcolor = blue,
            urlcolor  = blue,
            citecolor = blue,
            anchorcolor = blue]{hyperref}

\usepackage{todonotes}

\makeatletter
\renewcommand\paragraph{\@startsection{paragraph}{4}{\z@}%
            {-2.5ex\@plus -1ex \@minus -.25ex}%
            {1.25ex \@plus .25ex}%
            {\normalfont\normalsize\bfseries}}
\makeatother

\newcommand\pubnumber{Preprint}
\newcommand\pubdate{\today}

\def\Title#1{\begin{center} {\LARGE #1 } \end{center}}
\def\Author#1{\begin{center}{ \sc #1} \end{center}}
\def\Address#1{\begin{center}{ \it #1} \end{center}}

\newcommand\pubblock{\rightline{\begin{tabular}{l} \pubnumber\\
         \pubdate \end{tabular}}}
\newenvironment{Abstract}{\begin{quotation} \begin{center}
                       ABSTRACT
     \end{center}\bigskip  }{\end{quotation}}

\newcommand\snowmass{\begin{center}\rule[-0.2in]{\hsize}{0.01in}\\\rule{\hsize}{0.01in}\\
\vskip 0.1in Submitted to the  Proceedings of the US Community Study\\ 
on the Future of Particle Physics (Snowmass 2021)\\ 
\rule{\hsize}{0.01in}\\\rule[+0.2in]{\hsize}{0.01in} \end{center}}

\begin{document}

\pubblock

\Title{Summary Report\\Topical Group on Application and Industry\\
Community Engagement Frontier\\
Snowmass 2021\\
}

\bigskip 

\Author{Farah Fahim$^1$, Alex Murokh$^2$, Koji Yoshimura$^3$}
\Address{$^1$Fermi National Accelerator Laboratory, Batavia, IL, 60510, USA}
\Address{$^2$Radiabeam Technologies LLC, USA}
\Address{$^3$ Okayama University, Japan}
\medskip

\begin{Abstract}
  
\end{Abstract}

\snowmass
 
\clearpage

\section{Executive Summary
}   
\label{section:Section1}

HEP community leads and operates cutting-edge experiments for the DOE Office of Science which have challenging sensing, data processing, and computing requirements that far surpass typical industrial applications.  To make necessary progress in the energy, material, and fundamental sciences, development of novel technologies is often required to enable these advanced detector and accelerator programs. Our capabilities include efficient co-design, which is a prerequisite to enable the deployment of advanced techniques in a scientific setting where development spans from rapid prototyping to robust and reliable production scale.  This applies across the design spectrum from the low level fabrication techniques to the high level software development. It underpins the requirement for a holistic approach of innovation that accelerates the cycle of technology development and deployment. 

The challenges set by the next generation of experiments requires a collaborative approach between academia, industry and national labs. Just a single stakeholder will be unable to deliver the technologies required for the success of the scientific goals.

Tools and techniques developed for High Energy Physics (HEP) research can accelerate scientific discovery more broadly across DOE Office of Science and other federal initiatives and also benefit industry applications.
\subsection{Key Questions}

\begin{itemize}

\item How to enable sustainable innovation ecosystem which includes National labs, universities and industry by creating a mutually beneficial environment?

\item How can technologies developed at National labs be exploited in applications beyond HEP? 

\item Can we co-develop with other applications by engaging stakeholders from outside HEP?

\item Can we enable innovators to become entrepreneurs? Can we find mechanisms to enable deep tech commercialization?

\end{itemize}

\subsection{Findings}

\begin{itemize}

\item Difficulties effectively engaging with industry:
Funding for SBIR programs enables labs and universities to work with small businesses. However, For hardware development the time and funding scale of these programs are insufficient. Some technical developments are better suited to be addressed by partnering with larger companies but no formal funding mechanisms are available to initiate collaborations.
Moreover, engaging with big business is mired with issues associated with long lead times for establishing Cooperative Research and Development Agreements (CRADA). Furthermore, there is a lack of funding mechanisms to initiate and establish a partnership. 

\item Industry is often unaware of technology development goals and requirements for HEP.

\item National labs purchase tools and licenses from industry without leveraging economies of scale. DOE ICPT - Integrated Contracting and Purchasing teams are not effectively 
leveraged to create two-tier agreements, whereby pricing can be set based on DOE usage scale and volume purchasing is flexibly determined by the national labs. Moreover, the terms and conditions of purchase are renegotiated by every national lab which is extremely time consuming with completely different results across the national labs.

\item No mapping from science to technology goals is an impediment to enable cross-cutting initiatives across Office of Science and other federal agencies: Science goals of different federal experiments and programs are broad without overlap. These science goals drive the underlying technology requirements which have similar development trajectories and roadmaps. Common technology development goals are not identified and not utilized to foster joint programs. Cross- office of science programs do not have joint initiatives to enable synergistic development.
Cross-agency fund transfer is difficult to execute. Sometimes setting up an Inter Entity Work Order (IEWO) to use facilities or services that are outside DOE for e.g. at DOD or NASA labs can take more than a year.

\end{itemize}

\subsection{Recommendations}
\begin{itemize}

\item CEF.AI.01 – Public – Private Partnerships (Industry – National labs – Academia)

Tools and techniques developed for HEP research in addition to accelerating scientific discovery will also benefit industry applications especially enabling a mutually beneficial ecosystem in emerging technology areas - Quantum, AI/ML, Microelectronics. We need to translate the science goals to technology goals. Create programs, consortiums, centers that bring the various stakeholders together. 
\item CEF.AI.02 – Develop technology roadmaps to highlight synergies and technology readiness levels

Academia, National Labs and Industry form a spectrum from foundational research, advanced instrumentation to mature production. Create common engineering goals and roadmaps to accelerate technology readiness ensuring interaction with all the stakeholders at various stages of development.
\item CEF.AI.03 – Establish cross-agency initiatives

Establish application driven/ technology development engagement programs which have synergies with other funding agencies – including across Office of Science and all federal agencies. Cross cutting programs enable translational technology research. Hence focus on fundamental technology rather than fundamental science
\item CEF.AI.04 – Deep tech transfer initiatives

Establish programs that enable deep tech commercialization with appropriate incentives, funding and time scales. Consolidate and/ or establish entrepreneurial programs for employees across the National lab system.
\item CEF.AI.05 – Collective all of DOE approach for engagement

Employ an all of DOE approach instead of silo individual lab for procurement of common industry tools, licenses and services. Set DOE wide common terms and conditions with the flexibility for each lab to make the technical choices specific to their program. negotiate low-cost research licenses for basic science developments.

\item CEF.AI.06: Prioritize and simplify high risk, high reward opportunities

In some technical areas (i.e., FLASH radiotherapy), the high impact technology incubation by HEP ecosystem can produce significant, and occasionally disruptive, benefits to the society, within a decade timeframe. In these scenarios, we recommend to prioritize and simplify access by all domestic stakeholders to HEP facilities, expertise, and resources.
\item CEF.AI.07: Direct industry programs

There is also a growing interest in the community to improve support to the domestic industrial vendors providing critical technological capabilities to the HEP ecosystem. We recommend that DOE takes a proactive approach in establishing critical technology needs, and work directly with the qualified vendors to maintain and develop critical industrial capabilities, relevant to these needs.

\end{itemize}

\section{Introduction}
\label{section:Section2}

\subsection{Particle physics challenges breed innovation}
 
Experimental particle physics pushes the bounds of scientific discovery by advancing sensing, computing and communication technologies to explore the universe with unprecedented spatial and temporal precision. The unique and challenging science questions demand truly innovative solutions, and those solutions are often relevant for applications in areas like communications, computation and climate science. Tools and techniques developed for High Energy Physics (HEP) research can accelerate scientific discovery more broadly and also benefit industry applications.
 
In HEP, we build and operate the most complex experiments in the world including  collider detectors composed of more than 1 billion sensors and deep underground neutrino detectors holding 70 kilotons of liquid argon. With these ambitious experiments we aim to develop a complete understanding of dark matter, the Higgs boson, and the matter/antimatter asymmetry of the universe. These experiments set the challenges and opportunities for areas such as Microelectronics, Artificial Intelligence/ Machine Learning, Quantum development for HEP applications. 
 
Cutting edge particle detectors and accelerators create massive amounts of data which require powerful and energy efficient processing. The engineering design requirements for these detectors exceed those associated with industry including the Internet of Things for Industry 4.0, Smart cities, and Smart sensors for autonomous driving:
\begin{itemize}
     \item The data generated per second in just one large collider physics experiment is equivalent to the average internet traffic across North America.
     \item Experiments require more than one billion individual sensors with edge computing and ultra-low power and low-latency communication. The time scale to make decisions are a few orders of magnitude faster than typically required for industry applications.
\end{itemize}
New tools and fabrication techniques are required to build experiments that need to be  robust and  operate in extreme environments with long lifetimes:
\begin{itemize}
     \item The high radiation environment of a collider detector (1000x outer-space) demands development of techniques to radiation-harden commercial microelectronics.
    \item The technical challenge, cost, and environmental impact of powering and cooling one-billion sensors of a HEP experiment necessitates optimized devices with ultra low power consumption.

    \item Cryogenic operation (100 mK to 77K / -459F to -321F) of devices for readout of quantum sensors and cryogenic detectors necessitates collaboration with industry to develop cryogenic models and improve device performance.
    \item The inaccessible location of extreme environments requires long-term reliability (2-3 decades) for robust operation.
\end{itemize} 

Rapid prototyping at scale requires us to evaluate competing technologies:

\begin{itemize}
    \item We are early adopters of technology allowing us to assess and increase technology readiness level resulting in accelerated lab to fab innovation
    \item Some of our sensor arrays are almost twice the area of a basketball court requiring small volume prototyping. This gives us invaluable statistical insight into device properties, and influences improvements in material growth and fabrication.
    \item Deployment of compact detectors with lower size weight and power (SWaP) is a driver for evaluating hybrid integration and advanced packaging solutions
\end{itemize}

\subsection{Particle physics community is a key contributor in the Industry ecosystem}
 
National Laboratory, industry and academia collaboration enables the development of robust techniques which allows the maturation of novel technologies
\begin{itemize}
    \item  The High Energy Physics community collaborates extensively to enable breakthroughs in instrumentation: Academia, National Labs and Industry form a spectrum from foundational research, advanced instrumentation to mature production. We identify, analyze and support the adoption of most promising solutions.
 
    \item  Multidisciplinary teams to enhance co-design: In the case of Quantum, AI and microelectronics multidisciplinary teams of hardware experts, AI/quantum and microelectronics software/ algorithm developers, and domain scientists are required to develop the technology; this is a strength of the DOE National lab ecosystem.
 
    \item  Innovation translation: Small volume prototyping of systems based on novel devices, innovative circuit solutions and integrated architectures would enable accelerated demonstration leading to industry spin-offs or rapid adoption.
 
    \item  Tools and techniques developed for HEP research in addition to accelerating scientific discovery will also benefit industry applications.  This can enable broader adoption and cross fertilization of ideas. Furthermore, developing more challenging benchmarks which are oriented towards science applications will continue to breed more innovation. 
    
\end{itemize}

\section{Programs Enabling Deep Technology Transfer from National Labs }   
\label{section:Section3}

This section is a summary of \cite{https://doi.org/10.48550/arxiv.2203.15128}. DOE National labs are engaged in cutting edge research for next generation technologies to enable lofty goals set by experiments for basic science discovery. The next generation of DOE facilities for Colliders, Neutrinos, Astrophysics, etc. not only require demonstration of new concepts but also necessitate at-scale prototyping. 

While on one extreme universities are focused on surpassing the state-of-art, industry on the other extreme is focused on quality and repeatability. In-between this continuum, DOE experiments provide the motivation to go beyond proof of concept and manufacture at mid-scale, which can eventually lead to establishing a path for commercial production. Manufacturing for experiments increases the technology readiness level as well as making the process robust and cost effective. Hence DOE experiments and applications are a key driver of lab to fab innovation. 

In order to maximize the technology transfer potential, it is important to create an ecosystem where the technology inventors can create and sustain spin-offs/startups. Adapting the technologies developed for basic science to successful commercial ventures is a long and arduous process. In this white paper we evaluate the opportunities that can be made available to foster and sustain technology transfer.

Upon evaluating the opportunities,  seven major recommendations are present for increasing partnerships and commercialization at HEP-centric laboratories:
\begin{itemize}
     \item Aligning inventor royalty distribution policy across DOE: 
     
Inventor royalty distribution policy varies across the DOE complex. A proposed consistent royalty distribution might entail 34\% to the inventor(s), 33\% to the inventor's division, and 33\% to the Laboratory. The royalty portion distributed to the Laboratory or Division might be used to support HEP opportunities, consortium
investments, rewards programs, technology maturation to further the technology transfer
mission, and innovation/entrepreneurship educational opportunities for Laboratory staff
     \item Engaging with partnership intermediaries to accelerate commercialization: 

The majority of US DOE national laboratories that support high-energy physics facilities are small multi-program or single purpose laboratories that have limited resources available to support Technology Transfer (TT) and commercialization activities. The resource allocations for TT staff are generally proportional to other administrative support functions at these laboratories. However, commercial impact and viability is unrelated to laboratory size. Disruptive technologies can be created regardless of lab funding, program portfolio diversity or technology transfer support staffing. 

A Partnership Intermediary (PI) is a non-profit entity with specialized skills that can assist federal agencies and laboratories in TT and commercialization functions. 

Pilot programs funded in the past by the Department of Energy Office of Technology Transitions (OTT) to evaluate how a PI could interface with high-energy physics funded national laboratories showed some promise in assessing technologies for market pull, marketing technologies which are innovated in high-energy physics research areas, and matchmaking technologies at national labs with entrepreneurs in private industry who are interested in taking high-energy physics innovations from the laboratory to the market.  These intermediaries can provide support to accelerate innovation and commercialization from laboratories.

     \item Early identification of dual-use innovations:

Working with technology transfer entities, scientists and engineers developing new technologies can identify “dual-use” application cases where inventions identified early can proactively develop market analysis and case studies to support leveraging and commercializing technologies in areas beyond their immediate use.   Many of the barriers to capturing innovation from high-energy physics technology areas come from limited funds to pursue patents effectively, which requires technology transfer and laboratory leadership to evaluate innovations on their return on investment (ROI) and pull from the marketplace.    Developing value propositions and market analysis upfront on innovations with identified potential applications beyond their original intended use can accelerate acceptance and impact in the marketplace.

     \item Increasing technology transfer educational opportunities targeted to HEP researchers:
     
Providing educational opportunities to ramp up to an I-Corps level of engagement for high-energy physics researchers would be a great opportunity to provide the building blocks to researchers to enable more engagement in capturing innovations.
 Discussions on the types of intellectual property (ex. patents, copyrights), rights afforded to researchers from their innovations, and the mechanisms to engage with industry to advance their technologies would provide valuable resources and perspectives. 
 
     \item Public-private partnerships for accelerating HEP innovations:

Specifically, in technology areas such as accelerators, US federal program managers  have proposed developing public-private partnerships to foster and support small and large technology businesses who collaborate with the laboratories and serve as commercialization partners for critical technologies developed as part of facilities and experiments in high-energy physics. These public-private partnerships could serve as both advocacy and economic development entities for high-energy physics derived technologies, as well as matchmakers which aid companies and laboratories in forming collaborations which lead to commercialization outcomes.

     \item Extending other transaction authority:

An Other Transaction (OT) is a special mechanism used by federal agencies for obtaining or advancing R\&D or prototypes; it is not a contract, grant, or cooperative agreement, and there is no statutory or regulatory definition of “Other Transaction.” OTA is valuable in cases where the government needs to obtain R\&D and prototypes from commercial sources, but the companies equipped to provide them are unwilling or unable to comply with the government’s procurement regulations. The government’s procurement regulations and certain procurement statutes do not apply to OTs; thus, OTA gives agencies the requisite flexibility to develop agreements tailored to a particular engagement. 

While the Energy Policy Act of 2005 granted OTA to DOE at the agency level, it failed to authorize the labs to use it.  OTA could be offered as a unique authority provided by DOE to the labs that can enable HEP to showcase a more effective model for technology transition— but drive it at the local laboratory level, where the interaction with industry is vital for success. 
OTA is an ideal mechanism to help labs better identify market needs and become more valuable to the private sector because it can establish a formalized relationship where both parties have “skin in the game” early on in the research process, so markets can be better understood for deployment of technology.

     \item Establishing an entrepreneurial leave program:

An Entrepreneurial Leave Program (ELP) allows employees to take a leave of absence or separation from the laboratory in order to start or join a new company. ELPs encourage startup activities by reducing the risks faced by the employee entrepreneur. Some elements of an ELP may include business preparation / training, a means for licensing laboratory IP, continuity of health benefits during leave, and a mechanism for returning to work. ELPs are not implemented consistently across the DOE complex; some laboratories have ELPs while others do not.     

\end{itemize}
\section{Technology transfer with Scaleups}   
\label{section:Section6}

Industry partners typically fall into 3 categories: Small business, Large multinational organizations and mid-size scaleups. 
Our primary mechanism to collaborate with small businesses is through Small Business Innovation and Research (SBIR) programs. It is relatively easy to start this activity, many Phase I awards are available. Continuing a sustained development leading to higher impact is however limited since the number of Phase II grants are fewer and the available funding is not sufficient for deep tech development beyond an initial prototype or proof of concept demonstrator.

On the other hand, executing Cooperative Research and Development Agreements (CRADA) to work with large multinational companies is lengthy process often exceeding a year of negotiations over IP rights. In the long term these are extremely fruitful engagements but the slow start and fewer opportunities have created a high barrier to execute.

The intermediate option of Scaleups is unexploited and overlooked, here are a few recommendations on how to identify and engage with Scaleups.

Mechanisms for identifying Scaleups:
\begin{itemize}
    \item Bottom-up approach

There are enough online databases now that provide details of startups that have raised funds - including grants. These databases also can be queried based on verticals, technology, industry, geography, etc. Example - crunchbase.com; dealroom.com

 \item Top-down approach

Venture capitalists (VC) invest in many companies based on return on capital. They have a vested interested that the company used cutting edge technology and IP from reputed labs. Building relationships with the VCs will provide access to multiple scaleups in their portfolio.

\end{itemize}

Mechanisms to engage with Scaleups:
\begin{itemize}
    \item Laboratory Discovery days:

National Laboratory business development (BD) efforts with larger companies involves hosting “discovery days”. Product leads and problem owners from companies are invited to visit the lab and have discussions with technical/domain experts. There is considerable effort to plans these visits. Preplanning - build the top of the funnel with companies with criteria that helps labs to filter. Develop relationship at the CxO level and heads of the divisions. And, finally figuring a time to plan the visit. The last part is the hardest to find a time that works for senior executives in the companies and experts at labs. The success measured in number of discovery days that convert to projects. A similar discovery day to host scaleups at the lab to deep dive into their technology roadmap needs to be adopted.

    \item Through a Venture capital Firm:

Determining the labs value proposition is very important. If the company can procure services from a commercial service provider instead of a lab, it would be their preferred option. Generally it is considered difficult to get technology out of the lab because it is hard to get an exclusive license. Labs find it is better to give exclusive license to a large corporation, but in case of a startup – there is a chance it might fail. Universities are more comfortable with exclusive startup licenses. Leveraging contacts at venture capital firms is extremely useful, since they generally meet a lot of associated service providers such as lawyers, accountants, government lobbyists, labs. Labs need to create a summary description of what they do, which can be shared with companies to see if they are interested. Reaching out to companies and arranging meetings is important – but first defining what labs can do and how it’s different from others is essential to begin the conversation. Having access to an expense account to obtain fast services is also extremely beneficial.

    \item Partner with universities:

Use the university alumni resources - Some VCs such as ARCH looks for strong scientific founders who can help their companies at early stages and drive getting an exclusive license from the institution, which is not as easy to get at a National Lab. University model which allows staff member to spend one day a week on external projects helps facilitate such work. It is usually beneficial if they have a large lab and students who can get involved.

\end{itemize}
\section{Big Industry Engagement to Benefit HEP: Microelectronics Support from Large CAD Companies}   
\label{section:Section4}
This section is a summary of \cite{https://doi.org/10.48550/arxiv.2203.08973}. The development of modern microelectronics is a highly sophisticated and complex endeavor. There are few companies that have the capabilities to take on this challenge that
requires a range of deep expertise in device and circuit performance and their limitations,
as well as sophisticated CAD-EDA tools. Design of ASICs for DOE extreme environments,
such as high ionization radiation or cryogenic temperatures, does not have a significant commercial market to engage large companies in developing the required solutions. Currently
DOE national labs with academic and other collaborators spearhead the development of
chip design for next generation instrumentation required by the DOE mission.
The major bottleneck is the access to low-cost, high-volume microelectronics CAD tools.
The increasing complexity of designing in smaller geometry nodes has led to complicated
and expensive licensing frameworks, often with one license being shared among multiple
(e.g., more than 10) designers. As the technology node scales, even small designs take
significantly long completion time and are harder to debug with limited licenses. Moreover,
the existing framework is not suitable for collaborative development especially in joint teams
composed of groups from national labs, international labs, universities, and small businesses.
Finally, legal clauses for standard IP access are independently renegotiated by each DOE
lab, resulting in significant delays and different outcomes.

A centralized business model and legal framework negotiated between DOE contracting and
CAD tool/IP vendors, with input from the national labs needs to be developed. This model
then becomes the basis for engaging and pre-negotiating overall costs and terms and conditions with the vendors participating in the program. Each laboratory and its collaborators
can then independently procure CAD tools and IP based on their individual requirements.
The two-step system helps us provide economies of scale for the DOE microelectronics program, while enabling the labs to select tools most applicable to their team.

\subsection{Recommendations based on meetings with CAD Companies}
DOE Office of High Energy Physics hosted initial meetings with major CAD and EDA tool companies including Ansys, Cadence, Google, Keysight, Siemens, and Synopsys in 2021. Business collaboration models and recommendations are presented and discussed. Here's a summary of major recommendations:
\begin{itemize}
     \item Consider the concept of a DOE Collaborative Innovation Hub scoped for cooperation across the team shared access to CAD/EDA tools, training, and support.
     \item Establish a dedicated cloud-based communal participation between academia, DOE national labs, and CAD/EDA companies.
     \item Leverage successful solution frameworks (e.g. DARPA Toolbox \cite{DarpaToolbox}, DARPA Innovation Package, Europractice IC Service, DOD Cloud Access Rights) through the efficiencies of shared access.
     \item Incorporate some aspects of CAD/EDA companies' academia policies for research projects at national labs, to create a new class of research licenses.
     \item Leverage the academic network and cultivate talents to advance and promote innovations in semiconductor technologies.
     \item The solutions need to keep intact the premise of CAD/EDA companies' contributions, with special arrangements for commercializing research results.
     \item Build an Ecosystem including the CAD/EDA tools, available technologies, vendor support, and business models. 
\end{itemize}

\subsection{Mutual Impacts between HEP and Microelectronics Industry}
The DOE microelectronics development is miniature compared to the commercial microelectronics industry. But the industrial needs for innovative microelectronics for extreme environments (e.g., cryogenic operation, high ionization radiation) and the associated risk mitigation continues to increase, where DOE HEP microelectronics development has spearheaded for the past years. Major driving forces include Quantum and AI applications. Innovations and proof of concepts originated from the national labs are benefiting the microelectronics industry. Specialized cryogenic and energy efficient microelectronics design and integration techniques are adopted by the industry. Facilities at the national labs are also of growing interest to commercial partners. 

To CAD-EDA companies, DOE microelectronics teams have been a reliable resource to provide feedback on novel applications of the advanced CAD-EDA tools needed by DOE HEP projects. The close collaboration of national labs with the academic network also cultivates talent to advance and promote the workforce for the microelectronics industry.

\section{Application-driven engagement with universities, synergies with other funding agencies}   
\label{section:Section5}

This section is a summary of \cite{https://doi.org/10.48550/arxiv.2203.14706}.The success of the HEP Laboratory-University interaction is commendable. However, Laboratory-University interaction have been historically limited to the physics departments of university partners. With the advent of automation, electronic instrumentation, and most notably data-driven scientific discovery in HEP, opportunities of collaboration between HEP laboratories and engineering departments, particularly those focusing on computation, electronics, and data sciences are becoming of utmost importance. The accelerators and detectors and experiments in the HEP community represent one of the most significant, longest running and most fruitful engineering efforts undertaken. However, the synergistic relationship that exists between the HEP Laboratories and the university physics departments simply does not exist with engineering schools in the same widespread and systematic manner. Interactions with engineering departments do happen occasionally, but, by and large, they are opportunistic. Unlike our colleagues in Europe, the United State High-Energy Physics community lacks the programmatic ability to sponsor engineering research at our universities, to place graduate students on engineering projects long-term, to request research into areas of significance to projects and to be able to influence thesis and dissertation topics.

\subsection{Application-driven Project Partnership}

Most productive partnerships start out with entities joining forces to achieve outcomes targeting a concrete problem. In that regard, application-driven engagement of HEP labs with universities will serve as a starting point. Ultimately, the intersection of the research interests between an HEP lab and a university should encompass a problem that the HEP lab considers of strategic importance. 
There are fundamental engineering challenges that are expected to occur in a large variety of future HEP projects. Many involve intimate integration of measurement and on-demand computation requiring advanced microelectronics technologies. Equally many will generate a deluge of data in need of real-time processing with intelligent and autonomous agents. Viewing these major technological parameters as pillars of HEP activities, it is easy to define a sustainable overarching technical synergy in the form of hardware/system/computation co-design that is bound to remain relevant within the limits of not just a one-off project, but a whole generation of new HEP endeavors. 

\begin{itemize}
 
\item Engaging with Engineering departments:
A few concrete steps might further help to start define a landscape that is easier to navigate by the HEP scientists and university engineering departments. We would benefit from more explicit representation of these partnerships in established listings of funding opportunities as a starter. It would help both sides of the communities to have access to visible and specialized partnership support that is labeled clearly as HEP-Engineering effort. For instance, there are special fellowships reserved for graduate students from Physics to join an HEP lab as a yearlong fellow, while such specialty programs do not exist for Engineering graduate students in an exclusive manner. As for funded research projects, we may want to consider a class of projects designated as science-engineering partnership, where the expectation of all project application in that category could be to include two dedicated components reserved for science and engineering, respectively.  In current practice, many lab-university joint projects proposed for funding to DoE or NSF certainly contain aspects of both domains, yet their co-presence is not something we acknowledge explicitly. Dedicating a track of projects where these components are required (similar to NSF projects where a research and an education component are requested to be described individually) would enhance the awareness in the community on the relative importance and value of the engineering contributions. This would help attract more university engineering groups to partner with HEP.
\item Collaboration across DOE

\item Cross-agency collaboration with NASA, DOD and other federal agencies

\end{itemize}

\subsection{Workforce Training}
Universities are the institutions where the US workforce for science and engineering is produced. A strong and competitive workforce stream with predictive supply trends is of utmost national value. HEP labs can make a conscious effort to becoming influential on the training of the next generation technologists through a number of mechanisms. These steps would in fact not only impact graduate level trainees, but also undergraduate students:
\begin{itemize}

    \item All competitive research universities have strong traditions of involving undergraduate students in research. HEP labs can have “first dibs” on these students through systematic mechanisms.  We have observed many individual success stories of HEP lab members mentoring undergraduate students. A possible pipeline could be as follows. HEP labs recruit undergrads for internships through partner universities $\rightarrow$ students are trained in the setups and topics the HEP lab prioritizes $\rightarrow$  successful undergraduate students are channeled to graduate programs across a network of partner universities $\rightarrow$  universities recruit these into engineering PhD programs $\rightarrow$  co-advising models are used to mentor these students by both an engineering professor and a HEP scientist $\rightarrow$  feed the students back into the HEP workforce.  [As a side note, the significance of a steady stream of HEP-experienced technologists who chose not to enter the HEP workforce, but instead join the general workforce should not be overlooked.  Research projects frequently require the collaboration of industry partners and HEP-trained engineers in the general workforce provide natural inroads for collaboration.]
    \item	Another component of the workforce is international students who join US universities for graduate studies. Universities should welcome input from HEP labs on recruiting the next generation graduate students from this cohort with interdisciplinary (physics, science, and engineering) background. 
    \item	A practical mechanism for involving HEP scientists in recruiting and mentoring efforts of the universities is through establishment of joint academic appointments. All research universities have some form of Adjunct position defined for esteemed members of government labs and industry, so that they can have a direct link to provide input and interact with academics. Such positions should be reserved within engineering department for more HEP community members. These HEP scientists could then also be thesis advisors and thesis committee members of students. We must note that there are existing mechanisms to include external members on thesis committees. However, being an Adjunct faculty provides more direct access to university resources and information delivered on a daily basis to other faculty.
    \item	Finally, active advocacy by HEP lab members for engineering students for awards, such as the URA award will be invaluable. Often times, engineering faculty and engineering PhD students are not aware of all opportunities that exist within the HEP and national lab ecosystems. Guidance from HEP scientists will help lift entry barriers for them. 
\end{itemize}

\section{Transformative Technology for FLASH Radiation Therapy}   
\label{section:Section7}

Radiation Therapy (RT) provides lifesaving treatment to millions of cancer patients every year. One can argue that today RT represents the highest societal impact practical application of the particle accelerators.  HEP support has historically played a major role in development of the accelerator science and technology, and RT emergence as the standard of care is clearly one of the most important community engagement success stories in HEP history.

\noindent
Recently, a fundamentally different paradigm for increasing the therapeutic index of radiation therapy has emerged, supported by preclinical research, and based on the FLASH radiation effect. FLASH radiation therapy (FLASH-RT) is an ultra-high dose-rate delivery of a therapeutic radiation dose within a fraction of a second. Experimental studies have shown that normal tissues seem to be universally spared at these high dose rates, whereas tumors are not. The dose delivery conditions are not yet fully characterized. Still, it is currently estimated that large doses ($\geq$10 Gy) delivered in 200 ms or less produce normal tissue sparing effects yet effectively kill tumor cells. There is a great opportunity, but also many technical challenges, for the accelerator community to create the required dose rates with novel and compact accelerators to ensure the safe delivery of FLASH radiation beams. HEP ecosystem includes the world’s most advanced accelerator infrastructure, expertise, and facilities, and making these resources available to FLASH R\&D can play critical role in elevating FLASH-RT to practice over the course of the next decade.

\noindent
Many preclinical and first clinical results indicate a dramatic reduction of toxicity response at FLASH-RT dose rates compared to conventional dose rates. The first human patient was treated with FLASH in 2019 (a patient with recurrent cutaneous T-cell lymphoma). FLASH-RT produced a complete response and was exceptionally tolerated even after multiple non-FLASH skin irradiations had produced significant radiation damage from both photons and electrons. 

\noindent
Most FLASH experiments and preclinical studies have been performed with electrons and only very few with protons. Currently, electron FLASH studies are performed using 4-6 MeV electron beams from modified clinical linacs and provide the strongest, most consistent preclinical evidence for the FLASH effect. Experimental high dose rate photon beams have been formed using synchrotron radiation and keV X-rays from a tube (very early FLASH-RT studies) with mixed results for the FLASH effect. The FLASH effect has also been observed with protons using shoot-through beams from clinical CW or iso-cyclotrons. In shoot-through beams, the beam is not energy degraded, so the proton energy ranges from 230-250 MeV, i.e., the highest available proton energy with these cyclotrons. Once energy degraders are introduced into the beam to create lower energy proton beams, FLASH intensities cannot be achieved. Synchrotrons, even the rapid cycling 15 Hz ion synchrotron being developed at BNL, cannot produce the intense ion beams required for a clinical application of FLASH-RT– and only a very small volume can be irradiated at the cycle time of the synchrotron. 

\noindent
To overcome these limitations, a considerable research and development in this area is essential to optimize and clinically realize the curative potential of FLASH-RT with different radiation modalities. Below we list some of the specific accelerator R\&D programs relevant to this area:

\begin{itemize}
\item The Advanced Compact Carbon Ion Linac (ACCIL) is a program initiated by the Argonne National Laboratory to develop up to 1 kHz repetition rate, compact proton linac capable to deliver FLASH-RT doses. 

\item Scaling Fixed Field Gradient Accelerators (FFGA) are synchro-cyclotron style proton accelerators, which can operate at high repetition rates and high currents consistent with FLASH needs; most of the current R\&D programs on scaling FFGAs are performed by Japanese research groups.

\item Non-scaling FFGAs are particularly well suited for accelerating other ion species (i.e., carbon), and there is a pilot facility under construction at the National Particle Beam Therapy Center (Waco, TX).

\item Laser-driven accelerators can deliver very large doses of protons or high energy electrons from a compact source (both scenarios are potentially of interest to the FLASH-RT). The bulk of US program is centered at the LBNL BELLA laboratory.

\item The pulsed power based linear induction accelerator (LIA) using a multilayered bremsstrahlung conversion target also represent very promising technology in meeting FLASH-RT requirements, and there is a pilot program underway at LLNL.

\item Multiple groups are also working to develop FLASH-capable X-ray systems, including the ROAD initiative by UCLA/RadiaBeam, and PHASER initiative by SLAC/Tibaray.

\item Finally, one potential application, which can take advantage of the recent interest by HEP community towards novel cold RF technology, is a compact cold-RF Very High Energy Electron (VHEE) radiotherapy system, with relevant R\&D programs initiated at SLAC and at CERN.

\end{itemize}

\noindent
More information on this exciting initiatives are described in the Snowmass white paper fully devoted to the FLASH-RT topic \cite{https://doi.org/10.48550/arxiv.2203.11047}.  Here we just note, that the authors of this study believe that with the strong level of support by the HEP community, including access to accelerator facilities and expertise, a tremendous and possibly disruptive progress can be made in FLASH-RT within the next decade.

\section{Nurturing the Industrial Accelerator Technology Base in the US}   
\label{section:Section8}

There is a widespread perception within the accelerator community that a transfer of accelerator technology to US Industry is not a high priority. It is not uncommon for the US high energy physics community to develop state-of-the-art particle accelerator technology, and later having to buy that same technology from abroad for domestic projects. In contrast, the high energy physics communities in Europe and Asia work to nurture their domestic industrial bases, and this asymmetry in technology transfer policy creates an uneven playing field when US firms attempt to compete overseas, while greatly benefits foreign companies competing to serve DOE funded projects in the US. 

\noindent
This resultant relative weakness of the accelerator technology industrial base in the US has many undesirable consequences, including increased costs and reduced availability of critical components required by the labs, excessive and often politically unstable US dependence on foreign sources, a geographically localized and socially narrowed recruitment base for the technical personnel participating in the accelerator projects, and reduced recognition by society of benefits associated with the government investments into accelerator science and technology.

\noindent
In the white paper prepared by the industrial community \cite{https://doi.org/10.48550/arxiv.2203.10377}  we established that industry plays an important role in the scientific community and society in general, and yet the US domestic industry serving the needs of the DOE accelerator facilities has been struggling to achieve prominence. The specific case studies include AES and STI Optronics, which pioneered SCRF and undulator technologies, respectively, but could not sustain their business models without long term programmatic support from the customers. In both scenarios, the companies initially successfully acquired expertise, equipment, experience, and motivated customers base, all at a great cost and through decade-long efforts. However, neither of these companies could get a sustainable support from the DOE ecosystem which they served, resulting in their eventual collapse, and a destruction of not only the capital, but also unique and critical expertise, capabilities, with the long-term ripple effects through the supply chain affecting broader HEP community to this day.

\noindent
If the goal is to nurture and sustain a vibrant and competitive accelerator technology domestic industrial base in the US, some regulatory changes are being suggested. First, the DOE funding mechanisms that already exist, could be better utilized towards this goal. The US Small Business Innovation Research (SBIR/STTR) program is a great asset to help small businesses develop new capabilities, and it is the envy of many other countries, but the DOE does little to nurture these small businesses across the “Valley of Death”. Case studies are discussed in white paper, of the hardware and software accelerator projects and initiatives developed through the SBIR program.  In the authors’ view, one improvement would be to more closely align the program technical topics to the future procurement needs of the labs, and encourage the labs benefiting from the SBIR funded work to maintain the momentum and work with the industry beyond the SBIR funded phase.

\noindent
Of course, SBIR program is not sufficient to address broader problem of lack of support for the specialized industrial vendors, and other directed programs should also be implemented. Recent decades saw a proliferation of National Laboratories based commercialization centers built around the technology transfer activities. Yet, few can report successes and, in general the idea of technology transfer through funding the commercialization activities by the labs in authors’ opinion is counterproductive for the purposes of building the industrial vendors base. We believe it would be more beneficial to deemphasize technology transfer as a means of supporting the labs, and emphasize knowledge transfer as a means of supporting motivated businesses to expand capabilities of interest to the DOE programs. Laboratories should welcome domestic industrial interest in use of their expert consultants, specialized equipment, and IP, to develop cutting edge and economically viable commercial solutions that eventually benefit the accelerator community, and also create high quality jobs in the private sector.

\noindent
Finally, it is recommended that DOE establishes a method to identify key technologies that will be needed in a decade time frame and create new channels of direct funding to the qualified industrial enterprises to develop expertise, infrastructure, and capacity to meet such needs. It is also equally important to be able to help sustaining the companies that have already achieved critical capabilities but are not able to sustain them without a minimum volume of recurrent orders. 

\noindent
The white paper also discussed a need to simplify some of the laboratory procurement practices, and likewise explore various creative ways for industry and laboratories to collaborate on the prototype developments that would minimize the risks and maximize return to both sides. The accelerator community should also promote programs that facilitate direct and open communication channels between laboratory engineering and technical staff with their industrial counterparts (there are many conferences for scientists to attend and share their experiences, but not so many venues are available to technicians and engineers whose skills are essential and irreplicable in our field).

\noindent
These and many other steps could improve the quality and outcome of industry participation in the US DOE accelerator laboratories complex. Without such programmatic changes, and without having a seat at the table, it would be difficult for the US industry to achieve and sustain the prominence and excellence required to serve the needs of the HEP accelerator community and beyond.  On the other hand, a proactive approach by DOE, laboratories, and industry in improving the current situation, can produce a remarkable turnaround within a decade.

\section{Conclusions}   
\label{section:Conclusions}

High energy physics experiments can benefit immensely from creating sustainable collaborations between academia, national labs and industry. In order to develop a mutually beneficial ecosystem, it is important to translate science goals to technology roadmaps which can help identify the different stages of collaborative development.

HEP experiments are ideally suited to enable lab to fab innovation, since foundational research produced by academia can be accelerated to mature manufacturing and production by demonstrating mid-size scaling. The scaling can provide statistical insights into the fabrication techniques and device performance. 

Finally, technology development for HEP can immensely benefit from multi agency cross-cutting initiatives across other federal programs.
\clearpage

\bibliographystyle{unsrt}
\bibliography{bibliography}

\end{document}